

\BeginBibliography
\ifflat
  \input refforwd.tex
\else
  \input refs/refforwd.tex
\fi

\HeaderNumber=0 
\DefaultHeader=1 
\lp Preface\par 
\rp Preface\par

\immediate\write16{Preface}
\MakeFirstPagesEntry{Preface}

\noindent\hbox{\twentyfourbf Preface}
\bigskip\bigskip
\noindent
This book is devoted to exact solutions of quantum field theory models
(in one space plus one time dimensions). We also study
two dimensional models
of classical statistical physics, which are naturally related to these
problems.
Complete descriptions of the
solvable model are given by the Bethe Ansatz which was discovered by
H.~Bethe in 1931 \bb{\bethe} while studying the Heisenberg
antiferromagnet.  The Bethe Ansatz has been very useful for the
solution of various problems \bb{\gaudin}, \bb{\izyumov},
\bb{\liebII}--\bb{\liebIV}, \bb{\mcguire}, \bb{\shastry}, and \bb{\yangI}.

Some of the Bethe Ansatz solvable models
 have direct
physical application. A famous problem solved by the
Bethe Ansatz is the Kondo problem (see \bb{\tsvelick} and~\bb{\andrei}).
Another model is the Hubbard model \bb{\liebI}
which has a relation with high temperature
superconductivity.
An important application of the
Bethe Ansatz is in nonlinear optics where cooperative spontaneous
emission of radiation can be described by an exactly solvable
quantum model \bb{\rupasov}.
Bethe Ansatz is very useful in modern theoretical physics
\bb{\polyakovI, \polyakovII}.
Correlation functions provide us with dynamical
information about the model. They are described in
detail in this book.

Bethe Ansatz solvable models are not free; they
 generalise free models of quantum field theory
in the following sense. Many-body dynamics of
free models can be reduced to one body dynamics.
With the Bethe Ansatz, many body dynamics can be reduced to
two body dynamics.
The many particle scattering matrix is equal to the
product of two particle ones.
This leads to the self consistency relation for the two particle scattering
matrix. It is the famous Yang-Baxter equation (a survey of articles
can be found in \bb{\jimboII})
which is the central concept of exactly solvable models.
The role of the Yang-Baxter equation goes beyond the theory of dynamical
systems. It is very important in the theory of both knots \bb{\yangII} and
quantum groups \bb{\jimboII}.

Quantum exactly solvable models are closely related to the theory of
completely integrable differential equations. The simplest relation is
provided by the quasiclassical limit.
Quantum correlation functions are described by classical differential
equations.
The modern way to
solve these equations, the Inverse Scattering Method, was
founded in 1967 by C.S.~Gardner, J.M.~Greene, M.D.~Kruskal,
and R.M.~Miura \bb{\gardner}. They studied the Korteweg-deVries equation.
P.~Lax showed that this equation can be represented as
a commutativity condition for two linear differential
operators \bb{\lax}.
It is interesting to note that the Lax representation is algebraically
related to the Yang-Baxter equation \bb{\izergin}.

The Inverse Scattering Meth\-od permitted  a
wide class of nonlinear differential equations to be solved.
The Toda lattice is one example (see \bb{\flaschkaI}, \bb{\flaschkaII}, and
\bb{\flaschkaIII}). These
equations have applications in various areas of physics:
plasma physics, nonlinear optics, nonlinear ocean waves, and others.
There exists a set of very interesting books on the Inverse Scattering
Meth\-od (\bb{\ablowitz}, \bb{\bulloughI}, \bb{\bulloughII},
\bb{\calogero},
\bb{\faddeevI}, \bb{\lamb}, \bb{\newell}, and \bb{\zakharovIII}).
One should also mention that there exist
multi-dimensional completely integrable differential equations. The most
famous is the self-dual Yang-Mills equation \bb{\atiyah}.
Other examples are the Davey-Stewartson \bb{\davey} and
Kadomtsev-Petviashvilli \bb{\kadomtsev} equations. Completely integrable
ordinary differential equations are
also extremely important such as the famous Painlev\'e transcendents
(see \bb{\itsII} and references therein). They also
arise in two dimensional gravitation  and
matrix models (see \bb{\brezin}, \bb{\douglas}, and
\bb{\gross}) and
in the description of quantum correlation functions \bb{\itsI},
\bb{\jimboI},
and \bb{\mcguire}.

Further development of the Inverse Scattering Meth\-od is related to
\bb{\zakharovII}
where the Hamiltonian interpretation was understood;
L.D.\ Faddeev and V.E.~Zakharov showed that the
solution of a model by the Inverse Scattering Me\-thod
can be considered as a transformation to action-angle
variables. This provides an opportunity for
quasiclassical quantization.
The quantum theory of solitons was constructed in
\bb{\dashenI}--\bb{\dashenIII}, \bb{\faddeevIII}, and \bb{\goldstone}
 where
it was shown that after quantization, solitons appear as elementary
particles in the spectrum of the Hamiltonian.

The Quantum Inverse Scattering Meth\-od was discovered in
\bb{\sklyaninI}--\bb{\sklyaninIII}.
It provides a unified point of view to the exact solution of
classical and quantum models.
The Quantum Inverse Scattering Meth\-od combines the ideas of the
Bethe Ansatz and the Inverse Scattering Meth\-od.
The first model to be solved by means of the Quantum
Inverse Scattering Meth\-od was the
Nonlinear Schr\"odinger equation
$$i \partial_t \Psi = - \partial_x^2 \Psi + 2c \Psi^\dagger \Psi \Psi.$$
The Lax representation for this equation  was constructed in
reference \bb{\zakharovI}.
The Bethe Ansatz for the quantum version of this equation
was constructed in \bb{\liebII} and~\bb{\liebIII}.
The Quantum Inverse Scattering method
permitted a reproduction of the Bethe Ansatz results starting
from the Lax representation. An important development of the
Quantum Inverse Scattering Meth\-od is related to
the study of differential equations for
quantum correlation functions.
It was shown in \bb{\itsI} that
differential equations for quantum correlation functions are simply related
to the original differential equation which was quantized.
The correct language for the description of correlation functions
is the language of $\tau$-functions. This is described in our book. We
should mention the papers \bb{\barouch} and \bb{\tracy} where
differential equations were first obtained for correlation functions
of Ising model.

The Quantum Inverse Scattering Meth\-od is a well developed
method (see reviews \bb{\faddeevII}, \bb{\jimboII}, \bb{\kulish},
\bb{\shastry},
and \bb{\thacker}).
It has allowed a wide class of nonlinear evolution
equations to be solved.  It explains the algebraic nature of the
Bethe Ansatz. Our book explains this method in great detail.

An important example solved by the Quantum Inverse Scattering Meth\-od
\bb{\sklyaninI}
is the sine-Gordon equation
$$\partial_t^2 u - \partial_x^2u + {{m^2} \over {\beta}} \sin\,\beta u =
0.$$
In relation with this model, we would like to mention the new book
by F.A.~Smirnov \bb{\smirnov}.
Algebraic Bethe Ansatz is related to
Quantum Groups \bb{\drinfeld}.
It is also deeply related to
Zamolodchikov's theory of factorized S-matrices \bb{\zamolod} and
to the theory of exactly solvable lattice models
in classical statistical physics
(the best review of these models is the book by
R.~Baxter \bb{\baxter}) and to conformal field theory \bb{\belavin} and
\bb{\itzykson}.

In our book, we try to illustrate general statements
for a few simple models. Our main example is the Nonlinear
Schr\"odinger equation (in the quantum case this is the model
of a one-di\-men\-sion\-al Bose gas with $\delta$-repulsion).
We consider also the sine-Gordon model, the Heisenberg
antiferromagnet, and the Hubbard model.

Our book is divided into four parts. The first part
explains the coordinate Bethe Ansatz. We evaluate the energy
and momentum of excitations and the
scattering matrix in the thermodynamic limit. Normally, the
ground state of the model is a Fermi sphere (or Dirac sea).
Thermodynamics of the model are constructed very explicitely.

The second part explores the Quantum Inverse Scattering Meth\-od and
the Algebraic Bethe Ansatz. Classification of exactly solvable
models is given there. The important concept  of
the quantum determinant is introduced (it is  related to the
antipode in quantum groups). The partition function of the six vertex model
is represented as a determinant
for the finite lattice with domain wall boundary conditions.
The Pauli principle for  one-dimensional interacting bosons is
also discussed. Lattice versions of continuous models are constructed
in such a way that the most important dynamical
characteristics are preserved.

The third and fourth parts describe the theory of correlation
functions.
In Part III quantum correlation functions are represented as
determinants of some integral operators (of very special form).
The third part starts with an algebraic study
of scalar products. For example, it is proved that the
square of the norm of the Bethe wave function is equal to
the determinant of very simple matrices. It can be obtained
by linearisation of periodic boundary conditions (in the
logarithmic form) near the solution.
Correlation functions are represented as determinants of
some special integral operators (Fredholm type).
In the fourth part, differential equations for
correlation functions are derived.
Asymptotics of correlation functions are explicitely evaluated.
First we represent correlation
functions as Fredholm determinants of some integral operator.
This integral operator has a very special structure
which permits us to consider it as a
Gel'fand-Levitan-Marchenko operator of some other differential equation.
The last equation drives the correlation functions.
Asymptotics of correlation functions are evaluated explicitly, even
time- and temperature-dependent correlation functions are evaluated.

Another approach to correlation functions is also discussed.
It is related to conformal quantum field theory
and permits the evaluation of long distance asymptotics
of correlation functions. In the gapless situation,
correlation functions decay as powers of the distance.
These powers are called the
critical exponents. They depend on all parameters of the model
and we evaluate them explicitly. The method of their evaluation is
related to finite size corrections.
The central charge of the Virasoro algebra describing
the asymptotics of our models is generally equal to one.

Each part of the book is divided into chapters. The aims
and goals of each part are explained in
the separate introductions.
At the end of each chapter, there is a conclusion which both
summarizes and contains bibliographic comments as well as
the list of references. The list of references is at the end of each
chapter.

There is double enumeration of the formulas in the book. The
first number corresponds to the section number  and the
second is the formula number in the section.
When formulas from different chapters are referred to, we
precede the equation number with the number of the chapter.
If we refer to a section from another chapter then the section number
will be preceded by the chapter number.
Theorems are numbered separately in each chapter. Sections
with asterisks can be omitted in the first reading.

This book was started in St.\ Petersburg where we benefitted from
discussions with L.D.\ Faddeev, A.R.\ Its, N.A.\ Slavnov,
E.K.\ Sklyanin, N.Yu.\ Reshetikhin,
L.A.\ Takh\-ta\-jan, V.E.\ Zakharov,
V.N.\ Popov, F.A.\ Smirnov and A.N.\ Kirillov.
The book was finished in Stony Brook. We greatly appreciate
the creative atmosphere of the Institute for Theoreti\-cal Phy\-sics
at Stony Brook. We benefitted from discussions with
C.N.\ Yang, V.\ Jones, B.\ McCoy,  B.\ Sutherland,
M. Fowler, H. Thacker,
H.\ Flaschka, A.\ Fokas, A.\ Newell, M.\ Ablowitz, J.\ Palmer, C.\
Tracy, L.L.\ Chau,
R. Shrock, W. Weisberger,
F. E{\ss}ler, H. Frahm, K. Schoutens,
F. Figueirido,  E. Williams, and S. Ray.
The authors
wish to thank David A. Coker for
proofreading and pedagogical suggestions that have
added to the clarity of the book.

We are grateful to NFS for grant PHY-9107261.
\InputBibliography
\vfill\eject
\endinput